\documentclass[fleqn,10pt]{wlscirep}
\usepackage[utf8]{inputenc}
\usepackage[T1]{fontenc}
\usepackage{textcomp}
\usepackage{layouts}
\usepackage{setspace}
\usepackage{longtable}
\usepackage[version=4]{mhchem}       
\usepackage{multirow}
\usepackage{array}
\usepackage{hyperref}
\usepackage[version=4]{mhchem}
\usepackage{float}

\usepackage{xr}
\usepackage{cleveref}

\definecolor{light-gray}{rgb}{0.9,0.9,0.9}

\usepackage{listings}

\titleformat{\chapter}[hang] 
  {\large\bfseries}
  {}
  {0em}
  {}

\titlespacing*{\chapter} 
  {0pt}    
  {-30pt}   
  {10pt}   

\newcolumntype{C}[1]{>{\centering\arraybackslash}p{#1}}

\newcolumntype{P}[1]{>{\centering\arraybackslash}p{#1}}

\title{Prediction of EDS Maps from 4DSTEM Diffraction Patterns Using Convolutional Neural Networks}

\author[1]{Mridul Kumar}
\author[1, *]{Yevgeny Rakita}

\affil[1]{Department of Materials Engineering, Ben-Gurion University of the Negev, Be'er Sheva, Israel}

\affil[*]{Corresponding author email: rakita@bgu.ac.il}


\begin{abstract}
Understanding the relationship between atomic structure (order) and chemical composition (chemistry) is critical for advancing materials science, yet traditional spectroscopic techniques can be slow and damaging to sensitive samples. Four-dimensional scanning transmission electron microscopy (4D-STEM) captures detailed diffraction patterns across scanned regions, providing rich structural information, while energy dispersive X-ray spectroscopy (EDS) offers complementary chemical data. In this work, we develop a machine learning framework that predicts EDS spectra directly from 4D-STEM diffraction patterns, reducing beam exposure and acquisition time. A convolutional neural network (CNN) accurately infers elemental compositions, particularly for elements with strong diffraction contrast or higher concentrations, such as Oxygen and Tellurium. Both extrapolation and interpolation strategies demonstrate consistent performance, with improved predictions when additional structural context is available. Visual and cross-correlation analyses confirm the model’s ability to capture global and local compositional trends. This approach establishes a data-driven pathway to non-destructive, high-throughput materials characterization.
\end{abstract}

\begin{document}

\flushbottom
\maketitle
%
%
\thispagestyle{empty}


\section*{Introduction}

Advancements in modern electron microscopy have enabled unprecedented insights into the structural and chemical properties of materials at the nanoscale. Among these techniques, four-dimensional scanning transmission electron microscopy (4D-STEM) has emerged as a powerful tool, providing rich diffraction information across scanned regions of a specimen \cite{10.1017/S1431927619001351,Savitzky2021py4DSTEM}. Complementing to 4DSTEM, energy-dispersive X-ray spectroscopy (EDS) has been a standard technique for chemical characterization, however, its reliance on high electron doses and extended acquisition times often leads to beam-induced damage to the sample, particularly in sensitive materials \cite{HODOROABA2020397}. Such as, some inorganic materials can observe the damage for electron beam energies as low as 1 keV \cite{schmiedParticleAnalysisSEM2002}.

While 4DSTEM records a diffraction pattern at every probe position and provides rich reciprocal space information, recent developments in machine learning (ML) have demonstrated remarkable capabilities in uncovering complex correlations between different types of experimental data and in accelerating traditionally time-consuming characterization methods \cite{https://doi.org/10.1002/inf2.12028}. In the context of electron microscopy, ML approaches have been applied to automate diffraction pattern analysis, denoise spectral data, and even infer material properties directly from imaging modalities \cite{zhang2020atomic,sadri2024unsupervised}. Building on these advances, we explore the use of ML models to predict EDS spectra from 4D-STEM diffraction patterns. Such an approach not only has the potential to significantly reduce beam exposure and acquisition time but also opens a pathway for high-throughput and non-destructive chemical mapping. By leveraging the synergy between diffraction and spectroscopic data, this work aims to establish a framework where machine learning serves as a bridge between structure and chemistry in electron microscopy.

Machine learning is a data-driven technique in which machines learn decision making based on data \cite{zhou2021machine}. It is transforming material science by enabling the faster and more efficient discovery, design, and optimization of materials. By analyzing large data sets from experiments, simulations, or the literature, machine learning models can predict key material properties such as conductivity, strength, or stability without the need for time-consuming laboratory work or expensive computations. It also aids in the automated analysis of microscopy and spectroscopy data, improving the speed and accuracy of material characterization. Moreover, machine learning helps optimize manufacturing processes by identifying ideal synthesis conditions and can even drive autonomous laboratories that learn and adjust in real time. In general, it accelerates innovation in fields such as energy storage, electronics, and structural materials \cite{wei2019machine}.
\\
One of the key areas where machine learning has demonstrated significant value is in the prediction of degradation phenomena such as corrosion, crack propagation, and fatigue in materials like steel, asphalt, and concrete \cite{agrawal2014exploration}. These capabilities make machine learning a powerful complementary tool for understanding and forecasting material behaviour under various conditions. When integrated with advanced characterization techniques such as 4D-STEM, machine learning can enhance the interpretation of complex datasets and uncover subtle patterns that may be overlooked using traditional analytical methods alone. This synergistic approach offers a promising pathway for gaining deeper insights into material properties and their evolution over time.

In this study, we extend this paradigm by demonstrating how a convolutional neural network (CNN) can be employed to infer elemental compositions directly from diffraction patterns data commonly acquired in 4D-STEM experiments. Our results show that the CNN effectively captures compositional trends, particularly for elements with strong diffraction contrast or higher concentrations, such as Oxygen and Tellurium. Both extrapolation (training on one iteration file and predicting on the rest) and interpolation (training on first and final iteration file) strategies yielded consistent performance, with noticeable improvements in interpolation cases due to the additional structural context provided. Visual comparisons and cross-correlation analyses further confirmed the model’s ability to reproduce not only the global distribution of elements but also local variations and inter-element relationships. These findings underscore the potential of ML-based approaches to complement and enhance traditional techniques like 4D-STEM, enabling more efficient and data-driven analysis of material properties at the nanoscale.

\section*{Results and Discussion}
\begin{figure}[!ht]
\includegraphics[width=\textwidth]{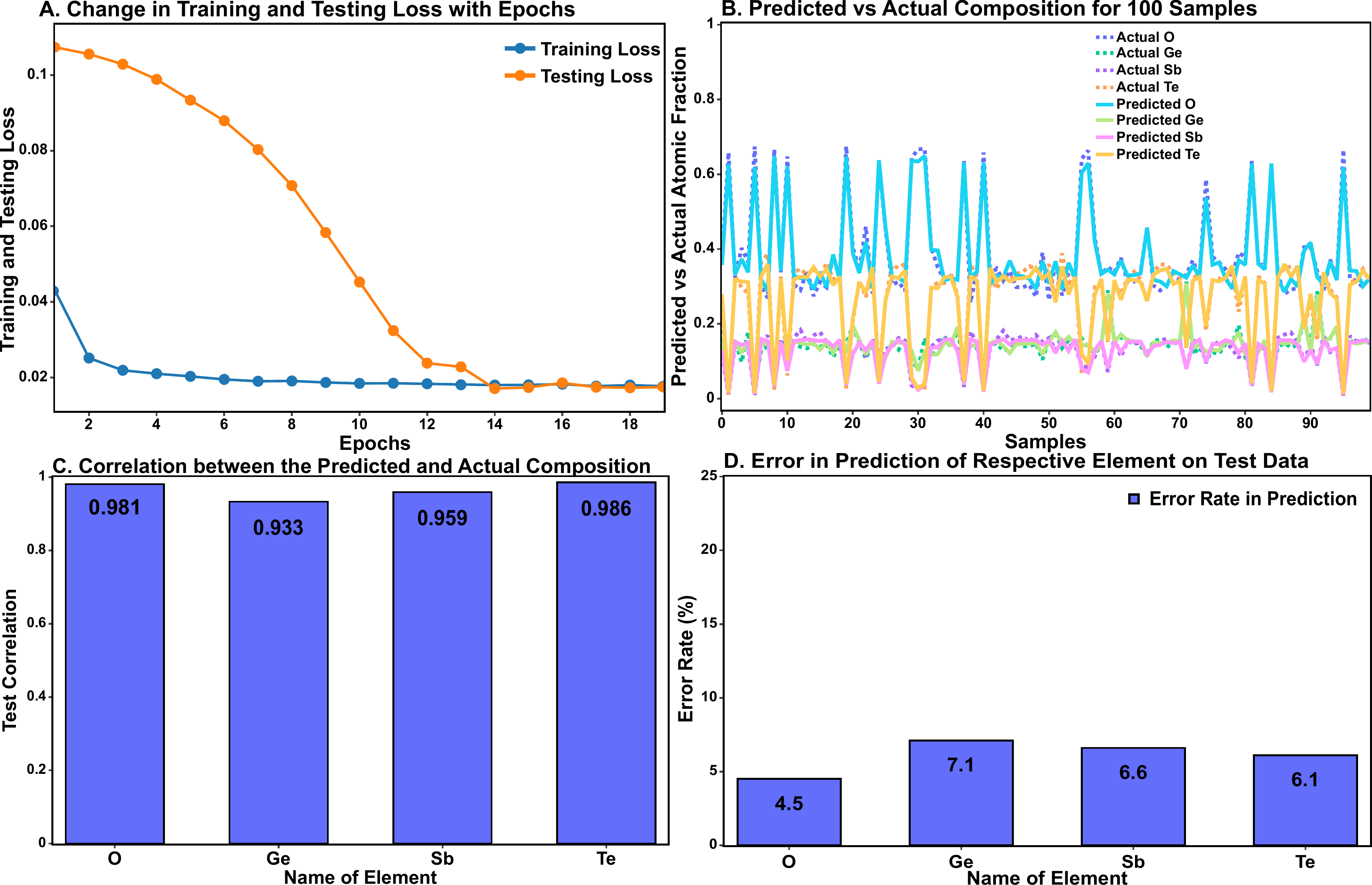}
\caption{A. Convergence of the CNN with the number of training epochs. B. Comparison between actual and predicted composition of the elements in the sample by our CNN on test data. C. Pearson correlation coefficient between predicted and actual composition of the elements in the sample. D. Error in the prediction of the composition in the sample by CNN. }
\label{fig:diff_to_eds}
\end{figure}

The Convolutional Neural Network (CNN) model was trained with the \lstinline|EarlyStopping| callback enabled, which halted the training process once the validation (testing) loss ceased to improve. This mechanism effectively prevented overfitting by ensuring that the model did not continue training beyond the point of optimal generalization. Overfitting occurs when a model learns noise or fluctuations in the training data instead of generalizable patterns, resulting in poor performance on unseen data \cite{zhou2021machine}. By monitoring the validation / testing loss, \lstinline|EarlyStopping| ensures that training is halted once the model's performance on unseen data plateaus, thus maintaining a balance between underfitting and overfitting. Although the initial improvement in validation loss was relatively slow, it eventually converged closely with the training loss towards the end of the training cycle (see Figure \ref{fig:diff_to_eds}A). The slower rate of improvement in validation loss compared to training loss is expected due to the model's initial focus on learning dominant patterns in the training data. As training progresses, the learned features begin to generalize better, leading to convergence between training and validation / testing losses in the later epochs.

To assess the model's predictive accuracy, the actual Energy Dispersive Spectroscopy (EDS) composition values were plotted against the predicted values (Figure \ref{fig:diff_to_eds}B). The close alignment between these values is further supported by strong Pearson correlation coefficients of 0.981, 0.950, 0.933, 0.959, and 0.986 for the elements O, Si, Ge, Sb, and Te, respectively (Figure \ref{fig:diff_to_eds}C).

The percentage prediction errors for O, Si, Ge, Sb, and Te were calculated as 4.5\%, 14.7\%, 7.1\%, 6.6\%, and 6.1\%, respectively (see Figure \ref{fig:diff_to_eds}D). Among these, Silicon (Si) exhibited the highest prediction error. The higher error percentage for silicon (Si) arises from its relatively small absolute presence in the sample. In such cases, even minimal deviations in the predicted values can lead to disproportionately high relative error percentages, which is a common challenge in low-abundance element prediction.

Overall, the CNN demonstrates robust performance in predicting EDS compositions, exhibiting strong correlation with ground truth values and acceptable error rates across all major constituent elements. However, the broader objective of this study is to develop a generalized machine learning model capable of predicting EDS composition maps directly from 4D-STEM diffraction patterns. To achieve this, the trained model must not only perform well on known data but also accurately predict EDS maps for entirely unseen 4D-STEM datasets. This requires the model to learn generalizable features that are independent of sample-specific variations, enabling reliable cross-sample predictions.

\subsection*{Training on One File, Predicting on Others}
To evaluate the generalization capabilities of the trained model, we adopted a cross file prediction strategy in which the model was trained on one file (0033) and tested on a separate file (0034 - 0038). This approach is particularly useful in assessing how well the model can capture underlying patterns that are not specific to a single file. During preprocessing, we observed that the sample exhibited oxidation near the bottom region, which could introduce bias or noise into the model (see Figure \ref{fig:blurred_vs_org_eds}). To mitigate this, we cropped each sample to retain only the central region of interest, resulting in final input dimensions of (70, 45, 256, 256) (see Code \ref{lst:read_cropped}). This selective cropping ensured that the model focused on representative, artifact-free areas of the data. The file-level separation, combined with careful preprocessing, enables a more accurate assessment of model robustness in practical applications.

\begin{lstlisting}[language=Python,caption={This code shows the reading of 4DSTEM file, subsequent cropping, application of filter and vectorisation of image data.},label={lst:read_cropped}]
def apply_filter(_filter,element,filters=["mean","gaussian","median","bilateral"]):
    #For application of the filter and vectorisation of image data.
    assert _filter in filters, f"Choose the correct filter from {filters}"
    if _filter == "mean":
        element = cv2.blur(image, ksize=(7, 7))
    elif _filter == "median":
        element = cv2.medianBlur(image, ksize=7)
    elif _filter == "bilateral":
        element = cv2.bilateralFilter(element, d=9, sigmaColor=75, sigmaSpace=75)
    elif _filter == "gaussian":
        element = cv2.GaussianBlur(element, (7, 7), 0)
    r, c = element.shape
    return element.reshape((r * c, 1))
def read_cropped_files(sample): #For reading and cropping
    current_filter = "gaussian"
    edspath = '/home/user/EDS/'
    diffpath = f"{sample} - 2 185 kx SI 4DSTEM.h5"
    X = py4DSTEM.read(diffpath).data / 2 ** 16
    X = X.reshape((X.shape[0]*X.shape[1],X.shape[2],X.shape[3]))[20*45:90*45,:,:]
    eds = []
    for element in ["O","Ge","Sb","Te","Au","Si"]:
    	filepath = f'{sample} - 2 185 kx SI-{element}-at.mrc'
    	file=mrcfile.read(os.path.join(edspath, filepath)
    	file = apply_filter(current_filter, file)
    	eds.append(file)
    y = np.concatenate(eds, axis = 1)[20 * 45:90 * 45, :]
    return X, y
\end{lstlisting}

CNN was trained and evaluated using two distinct strategies without modifying the architecture or hyperparameters from the previously optimized model (see Code \ref{lst:cnn}). This decision was based on the prior model’s demonstrated ability to converge well and perform reliably on similar diffraction-EDS datasets, making it a suitable candidate for further testing without additional tuning.

In the first approach, the model was trained using diffraction patterns and corresponding EDS compositions from the 0033 file. It was then used to predict the EDS compositions of the separate, unseen files 0034 through 0038 based only on their diffraction patterns (an extrapolation strategy). This method tested the model's ability to generalize to entirely new samples outside the training distribution. For better evaluation of the results generated from machine learning model, cross correlation between the predicted composition of the elements was calculated and it was then compared with the cross correlation values of the real composition of the elements (see Figure \ref{fig:cross_correl_ml_real}A). Predicted EDS compositions were also compared with the actual EDS composition for checking if the machine learning model is able to understand the patterns emerging in the diffractions and EDS composition (see Figure \ref{fig:cross_correl_ml_real}B).

In the second approach, the CNN was trained on combined data from files 0033 and 0038, incorporating both their diffraction patterns and EDS compositions. The trained model was then tasked with predicting the EDS composition of the intermediate files (0034 to 0037). This interpolation-based strategy aimed to evaluate the model’s capacity to capture and generalize spatial or compositional trends within a sequence of related samples. The basic idea was that the model would learn the changes if any from the initial and final file and would try to interpolate the compositional values in the intermediate files.

\subsubsection*{First Approach}
\begin{figure}[t]
\includegraphics[width=\textwidth]{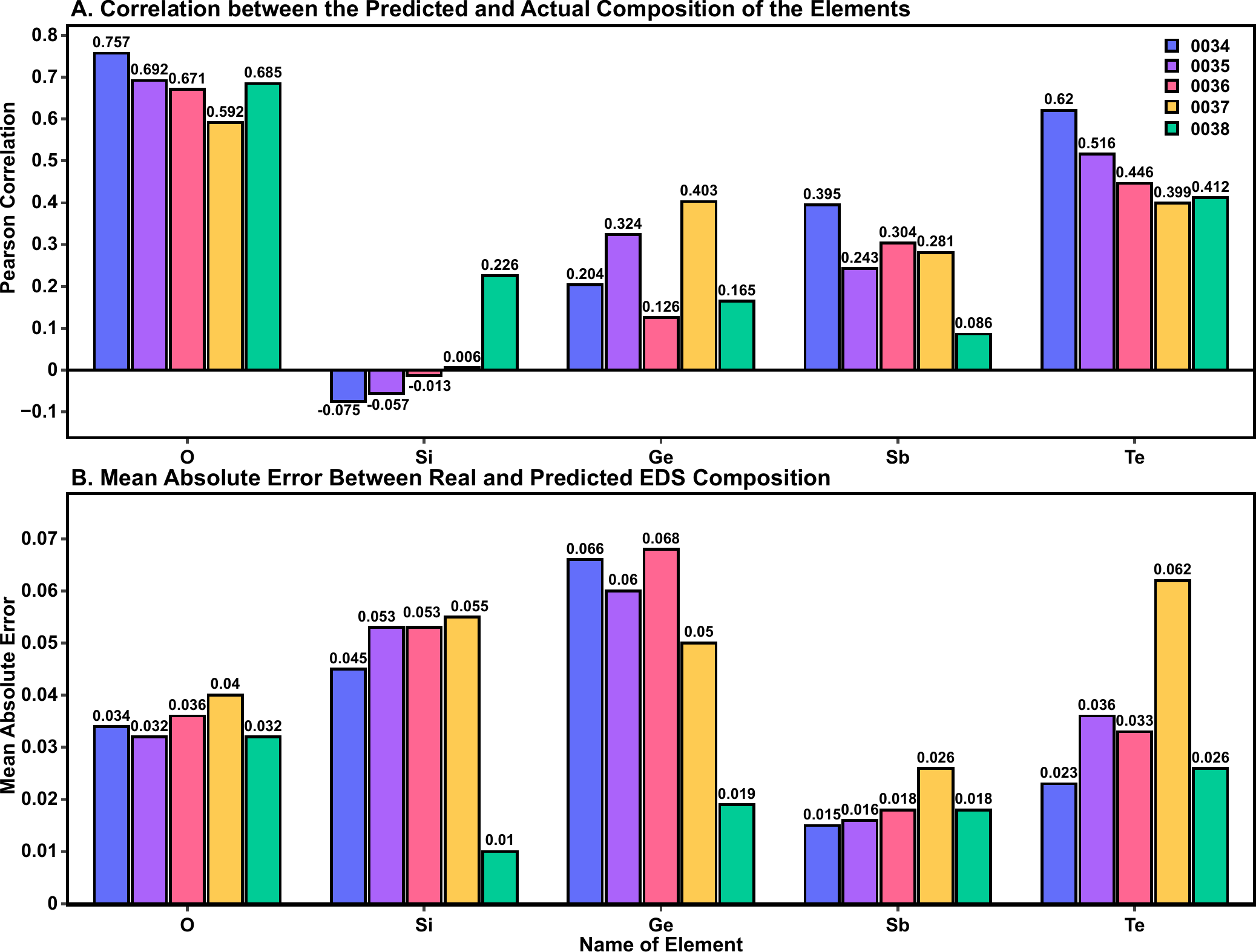}
\caption{A. Pearson correlation between the predicted and the real composition of the EDS values in the samples from 0034-0038. B. Mean absolute error calculated between the predicted and real composition values.}
\label{fig:corr_error}
\end{figure}

To quantitatively assess the CNN’s predictive capability, we computed the Pearson correlation coefficient and mean absolute error (MAE) between the predicted and ground truth EDS compositions for each element across all test samples. The correlation coefficient remained relatively consistent across samples, with mean values of 0.68 (O), 0.017 (Si), 0.24 (Ge), 0.26 (Sb), and 0.47 (Te), and corresponding standard deviations of 0.05, 0.10, 0.10, 0.10, and 0.08, respectively (see Figure \ref{fig:corr_error}A). These values indicate moderate to strong correlation for elements like oxygen and tellurium, while elements such as silicon and antimony showed weaker correlations.

In terms of MAE, the average deviations were 0.035 (O), 0.043 (Si), 0.052 (Ge), 0.018 (Sb), and 0.030 (Te), with standard deviations of 0.003, 0.017, 0.018, 0.004, and 0.014, respectively (see Figure \ref{fig:corr_error}B). The relatively higher standard deviation in both the correlation and MAE for silicon can be attributed to its low concentration in the samples. Even minor variations in prediction led to disproportionately large relative errors, reducing the statistical stability of its performance metrics.

These results suggest that the model performs more reliably on elements present in higher concentrations or exhibiting clearer diffraction features. Overall, the consistency of the correlation and low MAE values for key elements demonstrate the model's ability to infer compositional trends from diffraction patterns, with limitations observed primarily in elements with low concentrations in the samples.

\begin{figure}
\centering
\includegraphics[width=15cm]{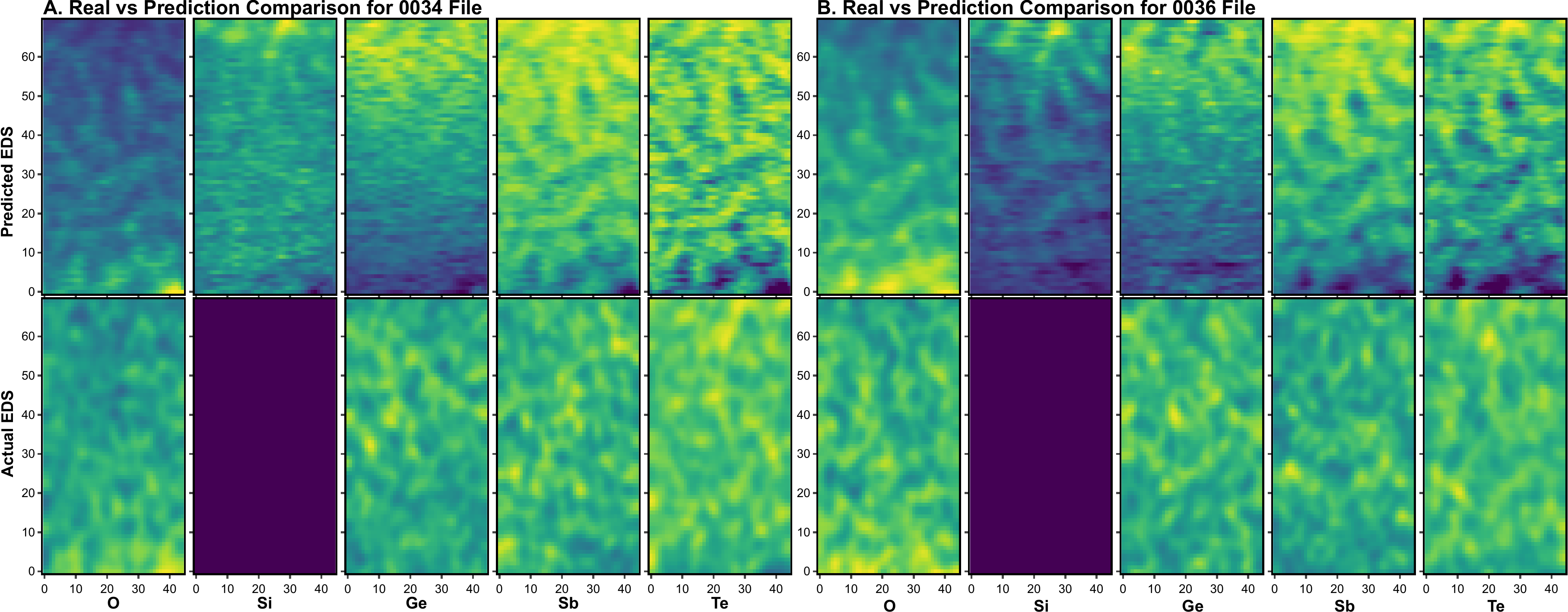}
\caption{This figure shows the comparison of the predicted and real EDS maps for files A. 0033 and B. 0036.}
\label{fig:real_pred_eds}
\end{figure}

Visual comparisons were also performed by plotting the predicted versus true EDS composition maps (see Figure \ref{fig:real_pred_eds}), enabling a qualitative assessment of how well the spatial distribution of elements was captured. Particular attention was given to compositional gradients, sharp interfaces, and spatially localized anomalies, as these features are critical for assessing the model’s practical applicability in real-world material characterization tasks.

In our observations, while the global compositional trends were generally preserved in the predictions, the model tended to underestimate local variations in the EDS maps. This smoothing effect suggests that the CNN may be biased toward learning broader spatial patterns rather than capturing fine-scale heterogeneities. Such behaviour may limit the model’s sensitivity to subtle structural or chemical inhomogeneities, particularly in elements present at low concentrations. Since, oxygen and tellurium are found in large composition, therefore, the model could also distinguish determine the local heterogeneities (see Figure \ref{fig:real_pred_eds}A and \ref{fig:real_pred_eds}B).

\begin{figure}
\includegraphics[width=\textwidth]{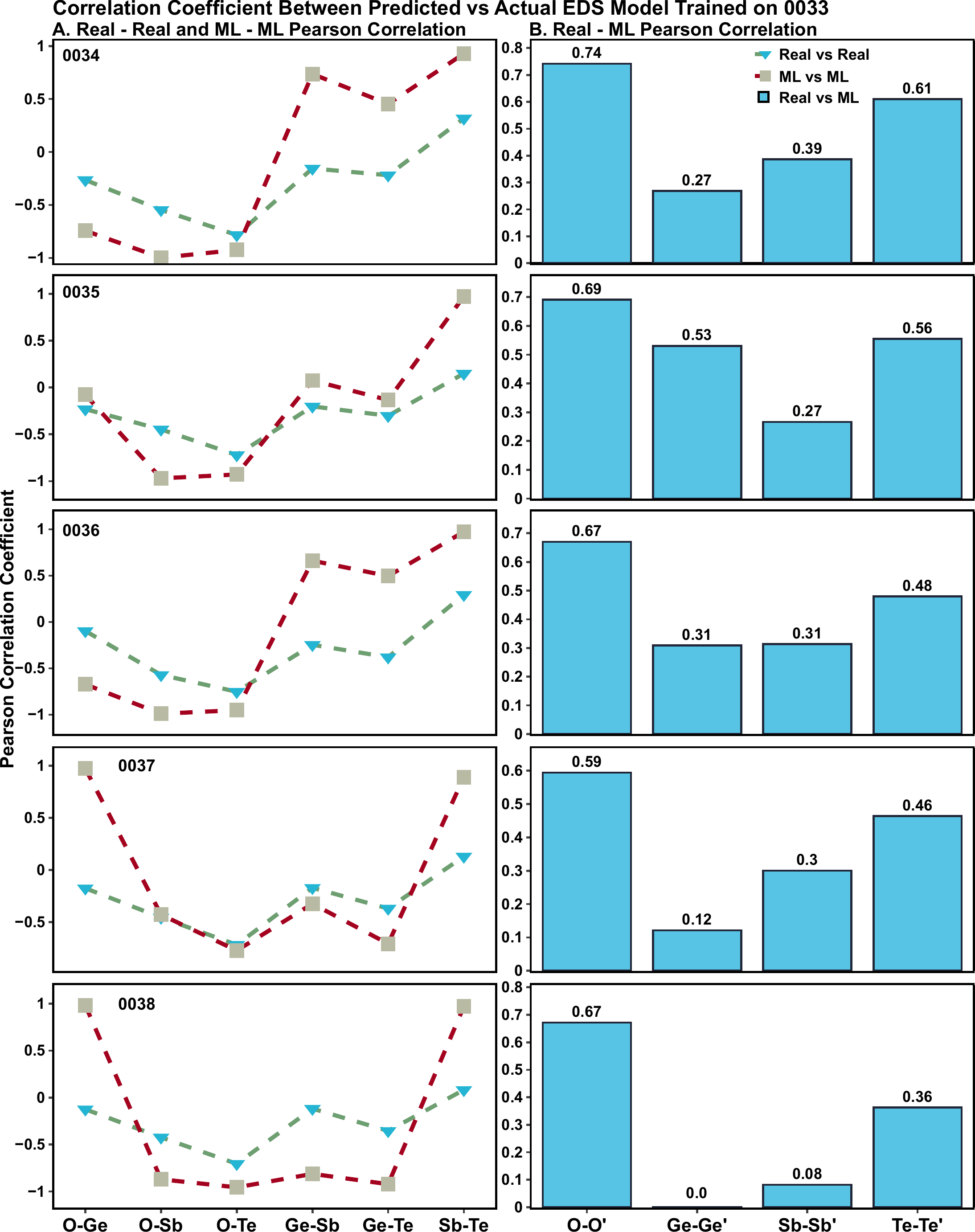}
\caption{Cross correlation trend between actual EDS composition of elements with each other and its comparison with cross correlation of predicted EDS composition of the respective elements with each other in first approach. Also the correlation of actual vs predicted EDS composition of elements.}
\label{fig:cross_correl_ml_real}
\end{figure}

To further investigate the trends in the predicted compositions, we computed and compared the cross-correlation matrices between elemental concentrations. Specifically, we plotted the cross-correlation between the concentrations of elements predicted by the CNN and compared it with the corresponding cross-correlation computed from the ground truth EDS data. The resulting correlation patterns revealed similar inter-element relationships in both predicted and actual compositions, indicating that the model successfully learned not only the individual elemental distributions but also the underlying compositional trends and dependencies among elements (see Figure \ref{fig:cross_correl_ml_real}A). This alignment reinforces the model’s ability to capture meaningful chemical relationships within the data. 

\begin{figure}
\includegraphics[width=\textwidth]{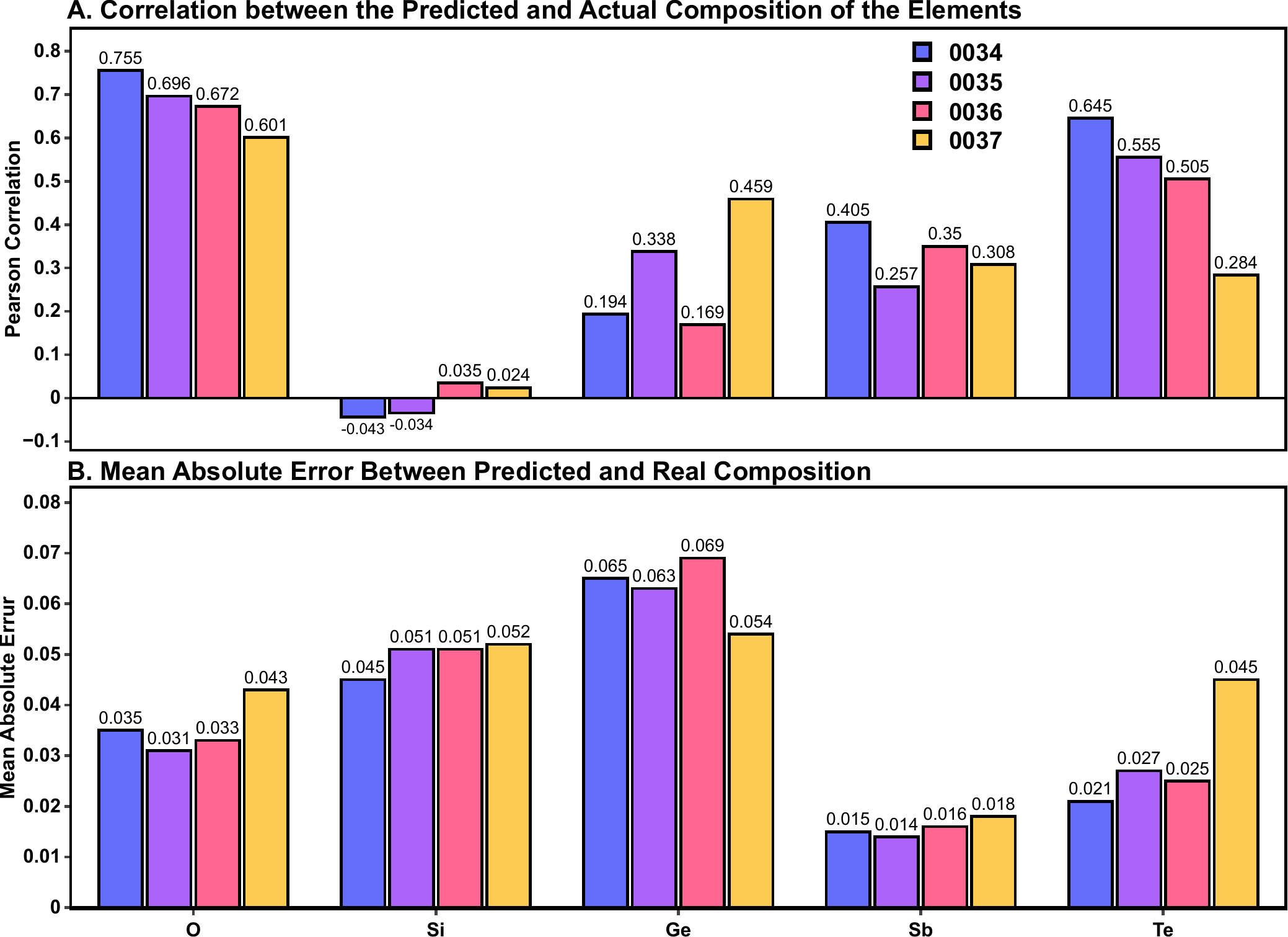}
\caption{A. Correlation between predicted and real composition of the elements in the sample. B. Mean absolute error between the predicted and real composition of the elements.}
\label{fig:corr_mae_2}
\end{figure}

\begin{figure}
\includegraphics[width=\textwidth]{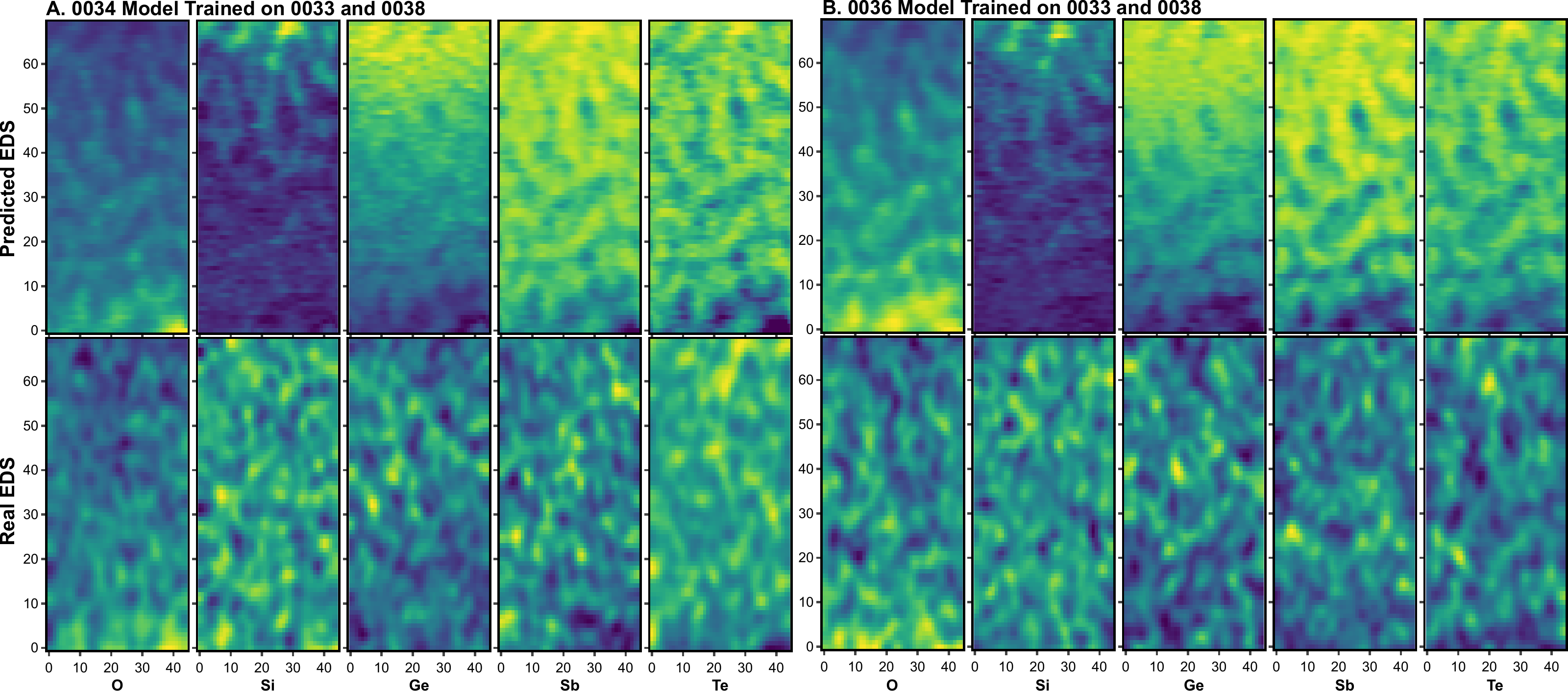}
\caption{This figure shows the comparison between the real and predicted EDS value of the files 0034 and 0036 for the interpolation approach in which CNN was trained on 0033 and 0038 files.}
\label{fig:real_pred_eds_2}
\end{figure}

\begin{figure}
\includegraphics[width=\textwidth]{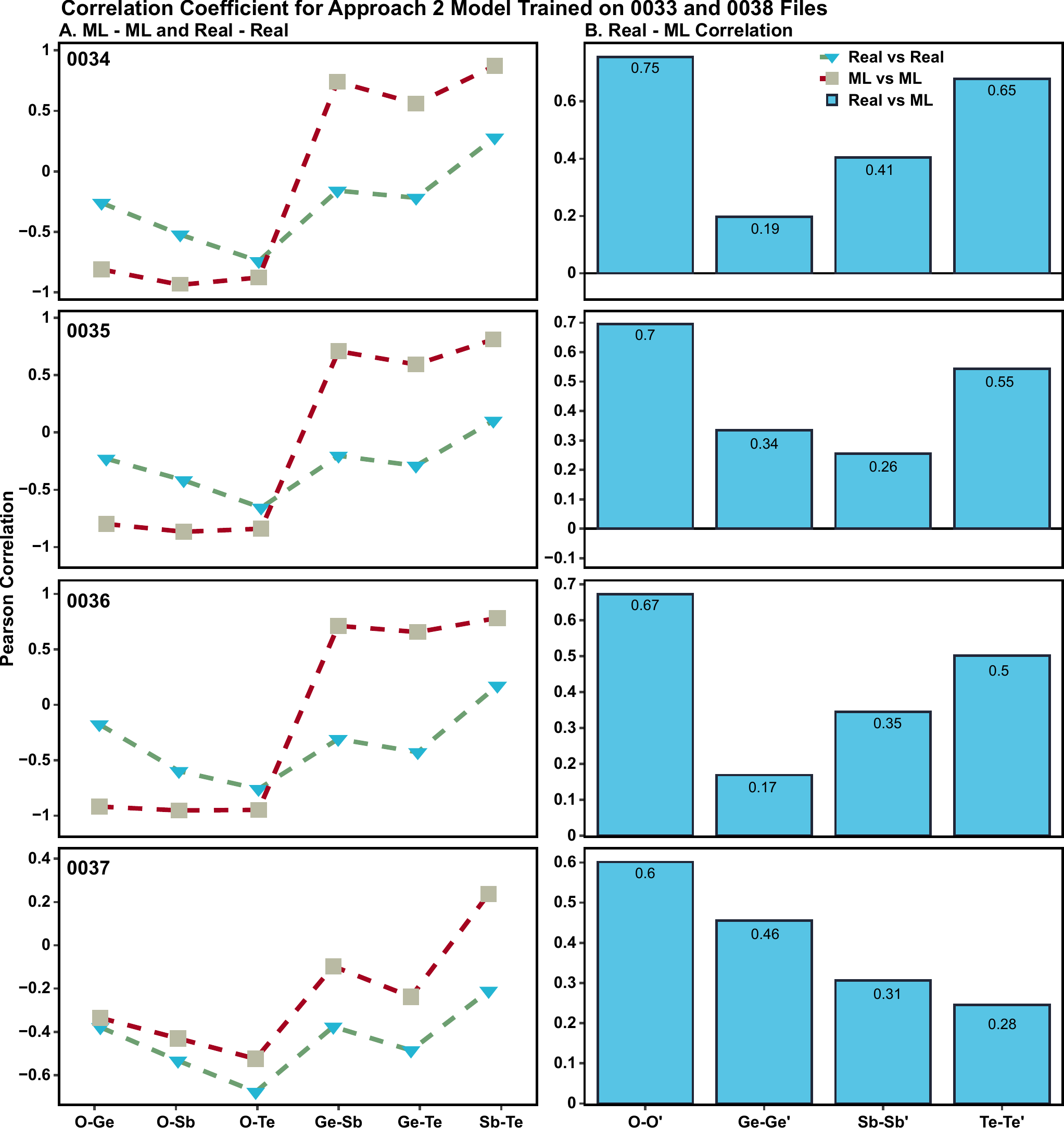}
\caption{This figure shows A. Comparison of cross correlation between predicted composition of the elements with the cross correlation of real composition of the elements in approach second. B. Correlation between real and predicted composition of the elements.}
\label{fig:ml_real_2}
\end{figure}

\subsubsection*{Second Approach}

In the second approach, rather than extrapolating to later files, an interpolation strategy was employed. The CNN model, previously used without any changes to architecture or hyperparameters, was trained on the diffraction patterns and EDS compositions from the endpoint files 0033 and 0038. The model was then used to predict the EDS compositions of the intermediate files, specifically 0034 through 0037.

To evaluate the model's performance in this interpolation setting, a methodology consistent with the previous (extrapolation) approach was followed. Initially, the Pearson correlation coefficient and mean absolute error (MAE) were computed between the predicted and experimentally measured EDS compositions. Following this, qualitative assessments were carried out through visual comparisons of elemental distribution maps. Lastly, to further quantify spatial consistency, cross-correlation analyses were conducted comparing predicted-predicted (ML–ML) and real-real EDS compositions for each element.

In our observations, the average Pearson correlation coefficients between the predicted and experimentally measured EDS compositions were 0.681 (O), –0.0045 (Si), 0.29 (Ge), 0.33 (Sb), and 0.497 (Te), with corresponding standard deviations of 0.055, 0.034, 0.117, 0.054, and 0.133, respectively (see Figure \ref{fig:corr_mae_2}A). These values indicate moderate to strong agreement for oxygen and tellurium, while the near-zero correlation for silicon again reflects poor predictive performance likely due to its low concentration and weak signal features in the dataset.

The mean absolute error (MAE) for the interpolation approach was found to be 0.0355 (O), 0.04975 (Si), 0.06275 (Ge), 0.01575 (Sb), and 0.0295 (Te), with standard deviations of 0.0046, 0.003, 0.005, 0.001, and 0.009, respectively (see Figure \ref{fig:corr_mae_2}B). Compared to the extrapolation results, a slight improvement was observed in the prediction accuracy for Ge, Sb, and Te. This suggests that training the model on both boundary conditions (files 0033 and 0038) provides additional context that enables the CNN to better interpolate intermediate compositions, particularly for elements with clearer compositional gradients.

For visual inspection of spatial similarity between the predicted and actual EDS maps, elemental composition maps for files 0034 and 0036 were plotted for both the predicted and ground truth data (see Figure \ref{fig:real_pred_eds_2}, for a better contrast the predicted data has been scaled). These visualizations provide qualitative insight into the model’s ability to reproduce compositional trends across the sample. Notably, the CNN successfully captured local variations in the concentration of oxygen and tellurium, which is consistent with the high Pearson correlation and low mean absolute error observed for these elements. This indicates that the model is not only learning global distribution patterns but is also sensitive to localized compositional changes, enhancing its utility for fine-resolution material characterization.

To further investigate the relationship between the predicted and actual elemental compositions, a cross-correlation analysis was performed, analogous to the approach used in the extrapolation strategy. Specifically, cross-correlation matrices were generated between the predicted and real compositions for each element (see Figure \ref{fig:ml_real_2}A and \ref{fig:ml_real_2}B). These matrices help evaluate not just point-wise accuracy, but the preservation of spatial and inter-element compositional patterns. The results from the interpolation approach closely resemble those obtained in the extrapolation setting, with noticeable improvements in correlation for certain elements, while others exhibited similar performance levels. This suggests that the model is able to retain structural relationships between elements across different prediction modes, further validating its generalization capabilities.

\subsection*{Improving the Accuracy of the Model}
Our initial convolutional neural network (CNN) model demonstrated that it could reliably capture the underlying correlations between 4D-STEM diffraction patterns and the corresponding EDS spectra (see Figure \ref{fig:real_pred_eds}). However, although the network reproduced the overall trends in compositional variation, it struggled to consistently predict the exact absolute values. Such systematic deviations are common in cross-modal prediction tasks, where the model learns relative relationships well but requires an additional calibration step for quantitative accuracy. Similar observations have been reported in spectroscopy and diffraction studies, where CNNs were effective at feature extraction but benefited from subsequent regression-based calibration to correct global biases \cite{wei2019machine}.

To address this, we moved towards a hybrid framework that integrates the non-linear feature-learning capacity of CNNs with post-processing methods such as linear regression, scaling transformations, or ensemble correction models. In this approach, the CNN provides a robust representation of the structural–compositional relationship, while regression aligns its predictions with the quantitative scale of EDS measurements. Hybrid strategies of this type have been shown to improve predictive accuracy in materials informatics tasks, for example by combining neural network outputs with regression calibration for molecular property prediction. To further strengthen the training data, high angle annular dark field (HAADF) was also included for the training of the CNN which was previously unused. For this purpose the output activation function was changed to Relu.

Let, $y_{CNN}$ denote the raw prediction from the CNN for a given input diffraction pattern and $y$ the corresponding ground-truth EDS intensity (or spectral component). A simple linear calibration model is defined as,

\begin{equation}
	y \equiv \alpha y_{CNN} + \beta
\end{equation}
Where, $\alpha$ and $\beta$ are the regression coefficients determined after the training of linear regressor. This correction ensures that the CNN retains its ability to capture non-linear structural compositional correlations, while the regression step enforces quantitative consistency with the experimental data. Depending on the distribution of residuals, more advanced schemes such as polynomial regression, quantile scaling, or ensemble-based corrections can also be incorporated.

\begin{figure}
\centering
\includegraphics[width=\textwidth]{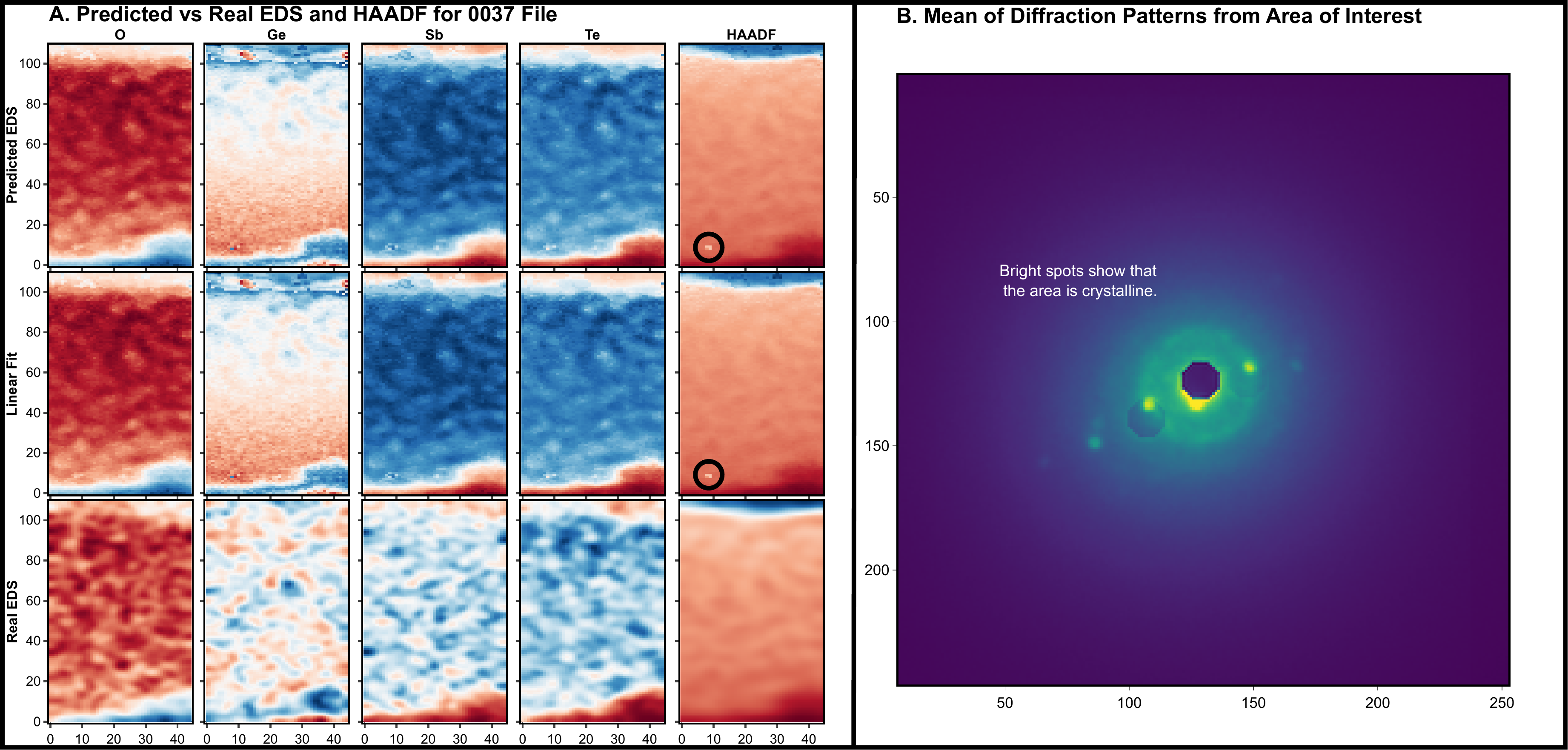}
\caption{A. Comparison between predicted EDS and HAADF by CNN and application of linear fit on the CNN output. It also compares these results to the real EDS maps of 0037 file. Interestingly, a small dot is observed (marked in circle) in predicted HAADF which is not visible in real HAADF. B. Mean diffraction pattern was calculated from the area of interest in A part. The dot turned out to be a crystalline region as apparent from the bright spots.}
\label{fig:eds_vs_real_lin}
\end{figure}

After training the convolutional neural network (CNN) using 4D-STEM diffraction patterns as input and both EDS and HAADF signals as outputs, we observed a substantial improvement in predictive performance. The mean absolute error (MAE) decreased significantly, reaching values as low as 0.0097. Closer inspection suggested that the EDS data alone contained a considerable amount of noise, which likely limited the model’s ability to converge to the true quantitative values. By including HAADF images relatively less noisy compared to EDS as an auxiliary training output, the model performance improved further. This multimodal approach enhanced feature extraction during training, resulting in stronger correlations between the predicted and ground-truth values of both EDS and HAADF signals.

Interestingly, in the case of dataset file 0037, the predicted HAADF output displayed a bright dot that was not present in the corresponding experimental HAADF image (Figure \ref{fig:eds_vs_real_lin}A). To investigate this anomaly, we examined the average diffraction pattern from the same region and found clear evidence of crystallinity, which manifested as a localized bright spot in the CNN prediction. In this case, the model incorrectly amplified the signal, yielding an anomalous white dot and a correspondingly elevated MAE (Figure \ref{fig:eds_vs_real_lin}B). This suggests that while the CNN was able to capture broad structural–compositional correlations, it was sensitive to local variations in diffraction contrast, particularly in regions with strong crystallinity. it can also be observed that the ML-predicted data is relatively lesser noisy as compared to the real EDS data.

\section*{Conclusion}
In this work, we demonstrated that CNNs can effectively capture the complex, non-linear relationships between 4D-STEM diffraction patterns and EDS spectra. While the raw CNN predictions accurately reflected relative variations, they exhibited systematic quantitative deviations from the experimental data. By introducing a regression-based calibration step, including HAADF as outputs, we reconciled the model predictions with ground-truth measurements, preserving the network’s ability to model structural and compositional correlations while achieving improved quantitative accuracy. Interestingly, even when the CNN failed to predict exact HAADF values, it exhibited higher mean absolute error (MAE), revealing the presence of crystalline regions. This highlights that, even in cases of imperfect predictions, machine learning models can provide new and meaningful insights.

This hybrid approach combining deep learning for feature extraction with simple regression for post hoc calibration offers a robust framework for translating qualitative trends into reliable quantitative predictions. Such strategies are broadly applicable in materials characterization and other domains where machine learning must bridge the gap between complex high-dimensional inputs and precise experimental measurements.

\section*{Materials and Methods}

\subsection*{Data Preparation}{\label{sec:data_prep}}
\begin{figure}[!t]
\centering
\includegraphics[width = \textwidth]{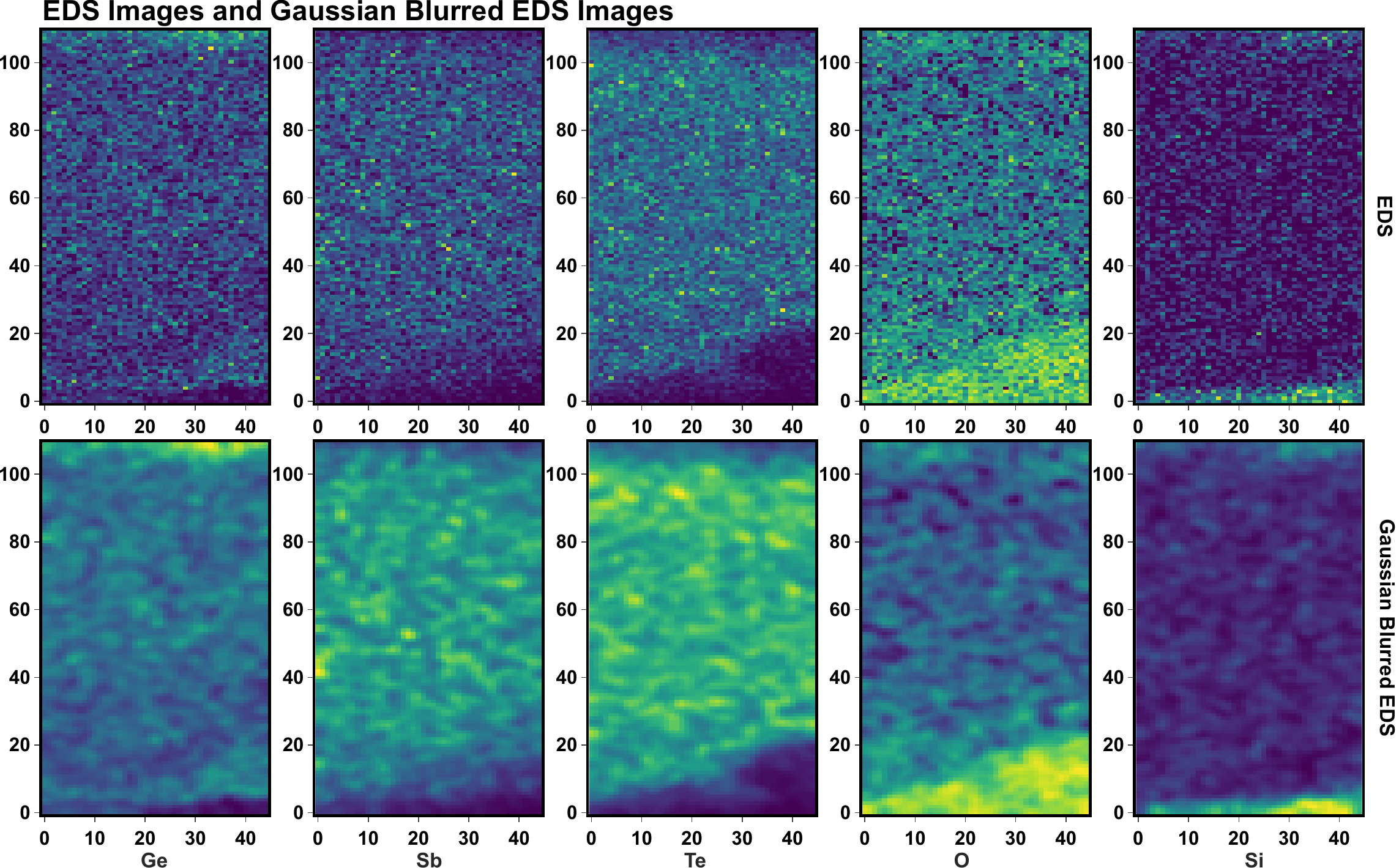}
\caption{Comparison between the original and Gaussian blurred EDS maps of all the elements in the sample 0033.}
\label{fig:blurred_vs_org_eds}
\end{figure}

A single sample of phase changing material (Ge-Sb-Te) was imaged by Thermo Fisher Spectra 200 STEM microscope. The sample was scanned multiple times at a magnification of 185k times to give mixed raster content (MRC) files namely from 0033 to 0038. Each diffraction pattern was $256 \times 256$ and real space dimension of the sample was $110 \times 45$. Other important information can be found in the Table \ref{tab:datainfo}.

\begin{table}[h]
\setlength{\tabcolsep}{2pt}
\renewcommand{\arraystretch}{1.2}

\caption{This table shows import properties of the data.}
\label{tab:datainfo}
\centering
\begin{tabular}{|C{6cm}|C{6cm}|}
\toprule

\textbf{Feature} & \textbf{Value} \\
\midrule

Spot Size & 2.5 nm \\

R Pixel Size & 1.053 nm \\

Q Pixel Size & 0.1451 $A^{-1}$ \\

Image Bit Depth & 16-bit \\
\bottomrule
\end{tabular}

\end{table}

\begin{figure}
\centering
\includegraphics[width=14cm]{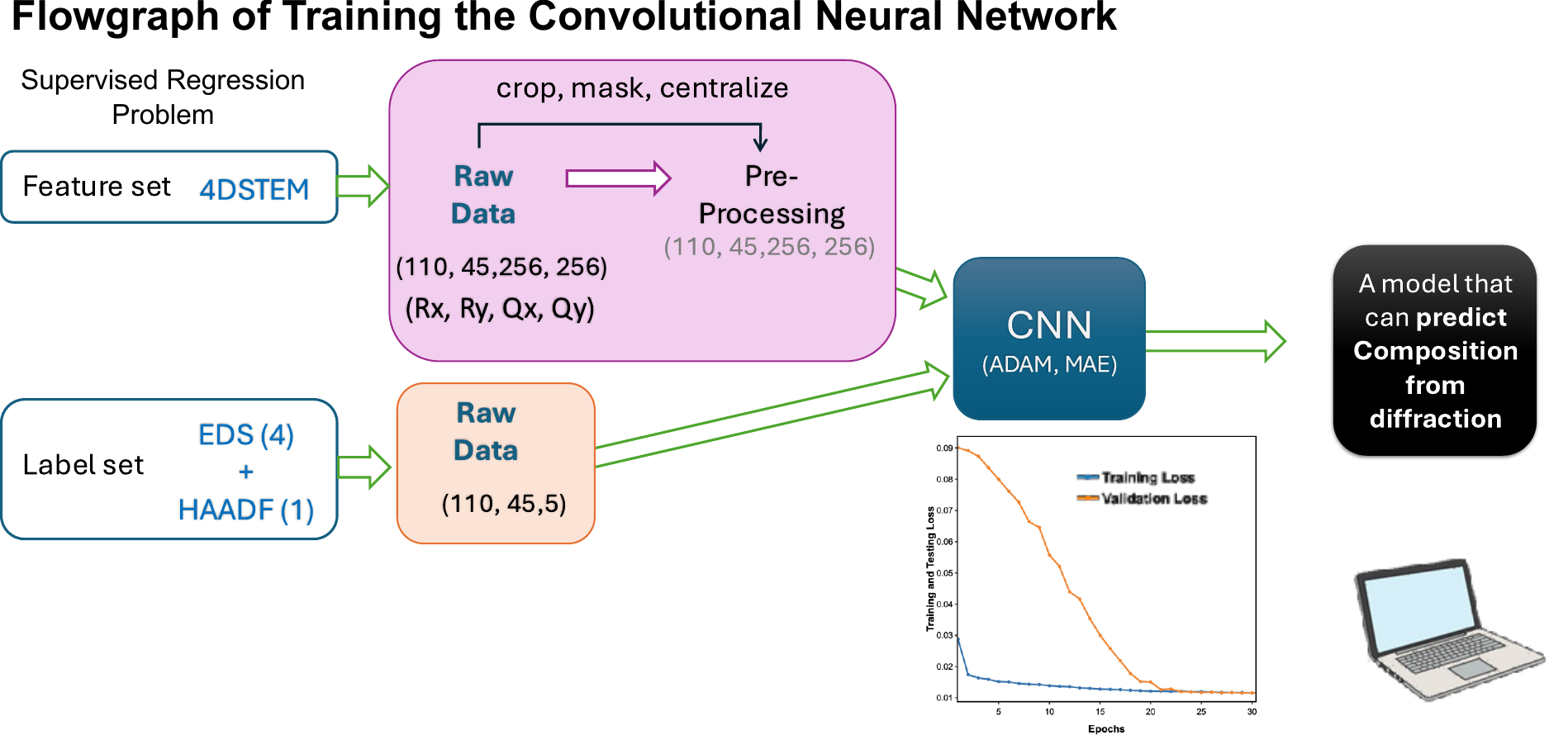}
\caption{This figure shows the steps involved in training the CNN such as preprocessing in which the 4DSTEM data was preprocessed with rlsconvert to mask, centralize and calibrate it. After that the data was reshaped and CNN was trained with Adaptive momentum (Adam) optimizer and MAE loss function.}
\label{fig:flowgraph}
\end{figure}

MRC files are 3D data files and we have to convert them to 4D hierarchical data format (HDF5) files, we used \lstinline|rlsconvert| package for this purpose. This package also helps in further processing such as removal of central beam, calibration, centralization of the diffraction patterns etc (see Figure \ref{fig:flowgraph}). 

The EDS composition maps produced are inherently grainy, therefore, requiring filtering techniques to smooth and enhance them for improved clarity. There are multiple filtering techniques to smoothen out images such as mean box filter, median filtering, and Gaussian blur. In our case, we applied a Gaussian blur filter with a channel width of 7 using \lstinline|opencv-python| package (see Code \ref{lst:eds_preproc}) \cite{bradski2000opencv}.
\begin{lstlisting}[language=Python, caption={Reading and blurring of EDS composition maps as data preprocessing step.}, label = {lst:eds_preproc}]
#Sample code for reading and applying gaussian blur filter on EDS maps.
import mrcfile
import cv2
eds = mrcfile.read("GST - 0033-Ge-at.mrc")
eds_blur = cv2.GaussianBlur(eds, (7, 7), 0) #Blurred EDS map.
\end{lstlisting}
Figure \ref{fig:blurred_vs_org_eds} shows the comparison between before and after the application of Gaussian blur on EDS maps of all the elements (Ge, Sb, Te, O, and Si) in sample 0033.

Diffraction patterns are also preprocessed for the removal of central bright beam, calibration, and centralization for the purpose of training with a machine learning model. \lstinline|rlsconvert| was used.

Since, the dimension of the sample is $110 \times 45$, the total number of diffraction patterns is 4950.

\subsection*{Choosing Machine Learning Model}

Diffraction patterns that we get from 4DSTEM are equivalent to image data. Spatial position of each pixel has significant meaning. Therefore, we need a machine learning model which can learn from the spatial position of each pixel. \\
Convolutional neural networks (CNNs) are one of the machine learning models falling under the subcategory of deep learning models which are designed to automatically and adaptively learn spatial hierarchies of features, making them ideal for tasks like image classification, object detection, facial recognition etc \cite{wenTimeSeriesAnomaly2019}. The key concepts of these models can be seen in Table \ref{tab:cnns}

\begin{table}[ht]
\setlength{\tabcolsep}{15pt} 
\renewcommand{\arraystretch}{1.5} 
\caption{Important features of a convolutional neural network.}
\label{tab:cnns}
\centering
	\begin{tabular}{|c|c|}
		\hline
		\textbf{Feature} & \textbf{Meaning} \\ \hline
		\textbf{Convolution Layers} &  It extracts features such as edges, textures and shapes. \\        \hline
		\textbf{Activation Function} & Adds non-linearity to the model. \\ \hline
		\textbf{Pooling Layer}  & Downsamples the output of convolution layers.\\ \hline
		\textbf{Dense Layers} & After convolution data is flattened and passed to dense layer. \\ \hline
		\textbf{Dropout} & Reduces overfitting. \\ \hline
	\end{tabular}
\end{table}

After the preprocessing in section \ref{sec:data_prep}, we have to make sure that the values of each pixel in the diffraction pattern lies between 0 and 1. It makes sure that the calculations are not computationally costly.

For creating, training, and testing the CNN we used \lstinline|tensorflow-gpu| with \lstinline|keras| API. Code \ref{lst:cnn} shows the architecture of our CNN.

\begin{lstlisting}[language=Python, caption={This code shows the CNN and key hyperparameters used for training on the 4DSTEM data.}, label = {lst:cnn}]
#Creation of Convoltional Neural Network.
model = keras.Sequential()
model.add(keras.Input(shape=(256, 256, 1)))
model.add(keras.layers.Conv2D(filters=32, kernel_size=(3, 3)))
model.add(keras.layers.BatchNormalization())
model.add(keras.layers.ReLU())
model.add(keras.layers.MaxPooling2D(pool_size=(2, 2)))
model.add(keras.layers.Conv2D(filters=64, kernel_size=(3, 3), activation='relu'))
model.add(keras.layers.MaxPooling2D(pool_size=(2, 2)))
model.add(keras.layers.Conv2D(filters=128, kernel_size=(3, 3), activation='relu'))
model.add(keras.layers.MaxPooling2D(pool_size=(2, 2)))
model.add(keras.layers.Conv2D(filters=256, kernel_size=(3, 3), activation="relu"))
model.add(keras.layers.MaxPooling2D())
model.add(keras.layers.Conv2D(filters=512, kernel_size=(3, 3), activation="relu"))
model.add(keras.layers.MaxPooling2D())
model.add(keras.layers.Dropout(0.5))
model.add(keras.layers.Flatten())
model.add(keras.layers.Dense(512, activation='relu'))
model.add(keras.layers.Dense(256, activation="relu"))
model.add(keras.layers.Dense(128, activation="relu"))
model.add(keras.layers.Dense(5, activation='softmax'))
model.compile(optimizer=keras.optimizers.Adam(0.0001), loss='mae', metrics=['mae'])
model.summary()
\end{lstlisting}

\subsubsection*{Training of the CNN}

In this study, we trained our CNN model to perform image regression (prediction of EDS composition maps based on the diffraction patterns) using an 80:20 train-test split (3960 diffraction patterns for training and 990 for testing) to ensure sufficient data for both learning and evaluation. The model was optimized using the Adam optimizer and learning rate of 0.0001, which combines the advantages of adaptive learning rates and momentum to achieve efficient convergence. Mean Absolute Error (MAE) was employed as the loss function, providing a direct measure of prediction accuracy by calculating the average magnitude of the errors without considering their direction. This setup was selected to emphasize robustness against outliers and ensure interpretable error metrics during model evaluation.

Tensorflow callbacks (\lstinline|tensorboard|for recording and saving the training progress,\lstinline|EarlyStopping| to stop the training as soon as the loss ceases to improve, and \lstinline|CSVLogger| for logging the training results in csv files) were utilized to optimize model training and evaluation, while also automating the saving of training results (see Code \ref{lst:cnn_training}).

\begin{lstlisting}[language=Python, caption={This code segment shows the reading, preprocessing and use of callbacks for training the CNN.}, label = {lst:cnn_training}]
X = py4DSTEM.read("0033.h5").reshape((4950, 256, 256))
y = []
for element in ["O","Ge","Sb","Te","Si"]:
	eds = mrcfile.read(f"GST - 0033-{element}-at.mrc")
	eds_blur = cv2.GaussianBlur(eds, (7, 7), 0) #Blurred EDS map.
	y.append(eds_blur)
y = np.concatenate(y,axis=1)
X_train,X_test,y_train,y_test=train_test_split(X,y,test_size=0.2,random_state=42)
ep=100
log_dir = "logs/fit/" + datetime.datetime.now().strftime("%Y%m%d-%H%M%S")
tensorboard_cb = TensorBoard(log_dir=log_dir, histogram_freq=1)
name = datetime.datetime.now().strftime("%Y%m%d-%H%M%S")
history = model.fit(X_train, y_train, batch_size = 64,
                    validation_data = (X_test, y_test), epochs = ep,
                    callbacks = [tensorboard_cb,
                    EarlyStopping(patience = 5, monitor = "val_loss"), 
                    CSVLogger(name + ".csv")])
\end{lstlisting}

\bibliography{bibliography}
\end{document}